\documentclass{emulateapj}
\usepackage{amsmath}
\usepackage{epsfig}
\usepackage{natbib}

\begin{document}
\begin{abstract}

Several groups have recently computed the gravitational radiation
recoil produced by the merger of two spinning black holes.  The
results suggest that spin can be the dominant contributor to the kick,
with reported recoil speeds of hundreds to even thousands of
kilometers per second.  The parameter space of spin kicks is large,
however, and it is ultimately desirable to have a simple formula that
gives the approximate magnitude of the kick given a mass ratio, spin
magnitudes, and spin orientations.  As a step toward this goal, we
perform a systematic study of the recoil speeds from mergers of black
holes with mass ratio $q\equiv m_1/m_2=2/3$ and dimensionless spin
parameters of $a_1/m_1$ and $a_2/m_2$ equal to 0 or 0.2, with
directions aligned or anti-aligned with the orbital angular momentum.
We also run an equal-mass $a_1/m_1=-a_2/m_2=0.2$ case, and find good
agreement with previous results.  We find that, for currently reported
kicks from aligned or anti-aligned spins, a simple kick formula
inspired by post-Newtonian analyses can reproduce the numerical
results to better than $\sim$10\%.

\end{abstract}


\title{Modeling kicks from the merger of non-precessing black-hole binaries}
\author{John G. Baker, William~D.~Boggs \altaffilmark{1}, Joan Centrella, Bernard~J.~Kelly,
Sean T. McWilliams \altaffilmark{1}, M.~Coleman Miller \altaffilmark{2}, James~R. van~Meter}
\affil{Laboratory for Gravitational Astrophysics, 
NASA Goddard Space Flight
Center, Greenbelt, Maryland 20771}
\altaffiltext{1}{University of Maryland, Department of Physics, College
  Park, Maryland 20742-4111}
\altaffiltext{2}{University of Maryland, Department of Astronomy,
College Park, Maryland 20742-2421}
\maketitle

\section{Introduction}

In the past two years, numerical relativity has undergone a revolution
such that now multiple codes are able to stably evolve the last few
cycles of the inspiral, merger, and ringdown of two black holes
\citep{Bruegmann:2003aw,Pretorius:2005gq,
Campanelli:2005dd,Campanelli:2006gf,Baker:2006yw,Baker:2005vv}.  An
important astrophysical output of such simulations is the net recoil
due to asymmetric emission of gravitational radiation, because this
has major implications for the growth of supermassive black holes in
hierarchical merger scenarios
\citep{Mer04,BMQ04,Hai04,MQ04,YM04,Vol05,Lib05,MAS05} as well as the 
evolution of seed black holes and current-day intermediate-mass black
holes \citep{Tan00,MH02a,MH02b,MT02a,MT02b,MC04,GMH04,GMH06,OL06,OOR07}.  
Analytical calculations \citep{Per62,
Bek73,Fit83,FD84,RR89,W92,FHH04,BQW05,DG06} have now been
augmented with full numerical results for non-spinning
black holes with different mass ratios \citep{HSL06,Baker06,
Gonzalez06}, 
for black holes with equal masses and spins initially 
orthogonal to the orbital plane
\citep{Herrmann07,Koppitz07}, for black holes with equal
masses and spins initially parallel to the orbital
plane\citep{Gonzalez07,Campanelli:2007cg}, for black holes with equal
masses and spins initially oriented at some angle between the orbital
plane and the orbital angular momentum\citep{Herrmann:2007ex,Tichy},
and for black holes with unequal masses and spins initially either
parallel to the orbital plane or oriented at some angle between the
orbital plane and the orbital angular momentum
\citep{Campanelli07}.

The parameter space for kicks with spin is large, so for astrophysical
purposes it is important to have a simple parameterized formula for
the kick that can be included in simulations of N-body dynamics or
cluster and galaxy mergers. For non-spinning black holes, the classic
\cite{Fit83} formula $v\propto\eta^2(\delta m/M)$ does a
reasonable job (although not perfect; see \citealt{Gonzalez06} for
numerical results), where for black hole masses $m_1$ and $m_2\geq m_1$
we define $M=m_1+m_2$, $\delta m=m_2-m_1$, and $\eta=m_1m_2/M^2$.
Initial explorations of equal-mass spin kicks also show evidence of
simplicity, with fits linear in spin reproducing the results of
\cite{Herrmann07} and \cite{Koppitz07}.  This encourages us to explore
a more general class of kicks.

Spins that are aligned (prograde) with the orbital angular momentum
may be of particular interest because it has been argued that
interaction with accretion disks will tend to align spins during
``wet'' mergers \citep{Bogdanovic:2007hp}.  The only previous numerical
studies for a black hole with an aligned spin were carried out by
\cite{Herrmann07} and \cite{Koppitz07}, which considered only cases
where the spin of the second black hole was anti-aligned (retrograde)
with the angular momentum and the masses of the black holes were
equal.  Here we compute the kick speeds from a $q\equiv m_1/m_2=2/3$
mass-ratio set of mergers, with spin parameters of 0 or 0.2, and
directions either aligned (prograde) or anti-aligned (retrograde) with
the orbital angular momentum. The symmetry of the configuration
therefore guarantees that the kick direction is in the orbital plane.
We find a formula that matches all of our kick speeds, and those of
\cite{Herrmann07} and \cite{Koppitz07}, to within $\sim 10$\%. If the
in-plane kicks can be generalized straightforwardly to more general
orientations, and added to kicks perpendicular to the orbital plane,
there is the prospect of simple astrophysical modeling of the
gravitational rocket effect for arbitrary black hole mergers. In
\S~2 we describe our initial data and methodology. In \S~3 we present our results, and 
discuss the implications of these simulations.

\section{Initial Data and Methodology}

In the following, we use geometrized units where Newton's
gravitational constant $G$ and the speed of light $c$ are set to unity
so that all relevant quantities can be represented in terms of their
mass-scaling.  For example, 1 $M_\odot$ is equivalent to a distance
of $1.4771 \times 10^5 {\rm cm}$, and a time of $4.9272 \times 10^{-6}
{\rm s}$.  Accordingly we express distances in terms of $M$, the
initial (ADM) mass of the system.

We simulated inspiraling black-hole binaries of various mass ratios
and spins, with the same initial coordinate separation in each case.
In these cases our initial mass ratio approximated either $2/3$ or
unity. Our simulations were performed with our finite differencing
code {\tt Hahndol} \citep{Imbiriba:2004tp}, which solves a 3+1
formulation of Einstein's equations.  Adaptive mesh refinement and
most parallelization was handled by the software package {\tt
Paramesh} \citep{paramesh}.  For initial data we used the
\cite{Brandt97b} Cauchy surface for black hole punctures, as computed
by the second-order-accurate, multigrid elliptic solver {\tt AMRMG}
\citep{Brown:2004ma}. We evolved this data with the standard
Baumgarte-Shapiro-Shibata-Nakamura
\citep{Nakamura87,Shibata95,Baumgarte98,Imbiriba:2004tp} evolution
equations, modified only slightly with dissipation terms as in
\cite{Huebner99} and constraint-damping terms as in \cite{Duez:2004uh}.
Our gauge conditions were chosen according to the ``moving puncture''
method, as in \cite{vanMeter:2006vi}.  Time-integration was performed
with a fourth-order Runge-Kutta algorithm, and spatial differencing
with fourth-order-accurate mesh-adapted differencing
\citep{Baker:2005xe}.  Interpolation between refinement regions was
fifth-order-accurate.

To explore the parameter space of non-precessing spin configurations
with some mass-ratio dependence, we have performed simulations for
seven different data sets of black holes with unequal masses, as well
as one equal-mass case. The initial-data parameters for these eight
data sets are given in Table~\ref{tab:initial}. The spins in these
simulations were always orthogonal to the orbital plane. Our choice of
initial tangential momenta was informed by a quasi-circular
post-Newtonian (PN) approximation \citep{Damour:2000we}.

Numerically, we can only directly specify the ``puncture'' masses
$m_{1p}$ and $m_{2p}$ of the two holes; to determine each hole's
physical (horizon) mass $m$, we first locate its apparent horizon
(using an adapted version of the {\tt AHFinderDirect} code; see
\citealt{Thornburg:2003sf}), and then apply the
\cite{Christodoulou70} formula:
\begin{equation}
m^2 = m^2_{irr} + J^2/4m^2_{irr},
\end{equation}
where $J$ is the magnitude of the spin angular momentum of the hole,
$m_{irr} = \sqrt{A_{AH}/16\pi}$, and $A_{AH}$ is the area of the
apparent horizon.

The parameters relevant to our discussion -- the mass ratio $q \equiv
m_1/m_2$ and the dimensionless spin parameters $\hat{a}_1 \equiv
a_1/m_1$ and $\hat{a}_2 \equiv a_2/m_2$ -- are listed in the first
eight rows of Table~\ref{tab:results}. We have striven to maintain $q
= 2/3$ in all of our unequal-mass simulations.

\begin{deluxetable}{lrrrrrr}
\tablecolumns{7}
\tablewidth{0pt}

\tablecaption{Initial data parameters.  Runs are labeled ``EQ'' for equal mass and
``NE'' for unequal mass. $J_1$ and $J_2$ are the spin angular momenta
of the two holes, either aligned (positive) or anti-aligned (negative)
with the orbital angular momentum. $m_{p1}$ and $m_{p2}$ are the
directly specified puncture masses of the holes. $P$ is the initial
transverse momentum on each hole, while $L$ is the initial coordinate
separation of the punctures. \label{tab:initial}}
\tablehead{\colhead{Run} & \colhead{$J_1 (M^2)$} & \colhead{$J_2 (M^2)$} & \colhead{$m_{p1} (M)$} & \colhead{$m_{p2} (M)$} & \colhead{$P$ (M)} & \colhead{$L$ (M)}}
\startdata
 ${\rm NE}_{--}$ & -0.032 & -0.072 & 0.374 & 0.586 & 0.119 & 7.05 \\
 ${\rm NE}_{-+}$ & -0.032 &  0.072 & 0.374 & 0.586 & 0.119 & 7.05 \\
 ${\rm NE}_{0-}$ &  0.000 & -0.072 & 0.374 & 0.586 & 0.119 & 7.05 \\
 ${\rm NE}_{00}$ &  0.000 &  0.000 & 0.382 & 0.584 & 0.119 & 7.05 \\
 ${\rm NE}_{0+}$ &  0.000 &  0.072 & 0.374 & 0.586 & 0.119 & 7.05 \\
 ${\rm NE}_{+-}$ &  0.032 & -0.072 & 0.374 & 0.586 & 0.119 & 7.05 \\
 ${\rm NE}_{++}$ &  0.032 &  0.072 & 0.374 & 0.586 & 0.119 & 7.05 \\
 ${\rm EQ}_{+-}$ &  0.050 & -0.050 & 0.480 & 0.480 & 0.124 & 7.00 \\
\enddata
\end{deluxetable}

\begin{center} \begin{figure}
\includegraphics*[scale=0.35,angle=-90]{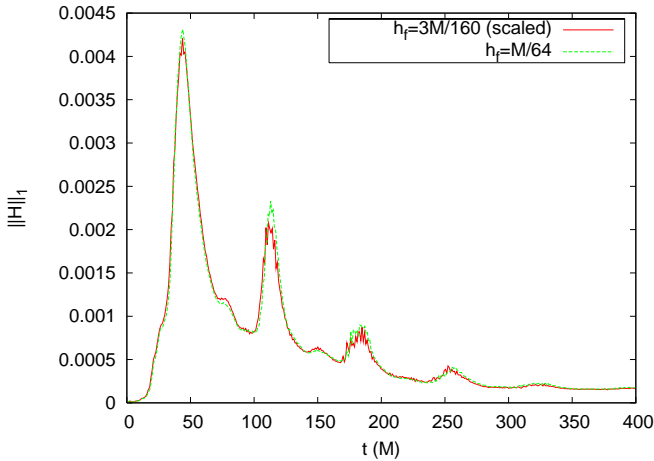}
\caption{L1 norm of the Hamiltonian constraint for the case of unequal masses with
prograde spins (${\rm NE}_{++}$) at two different resolutions,
computed on a domain extending to $|x_i|=48M$, which includes our
wave-extraction surface.  The finest level, containing nonconvergent
points near the puncture, has been excluded.  The lower resolution,
$h_f=3M/160$, has been scaled such that for third-order convergence it
should superpose with the $h_f=M/64$ curve.  }
\label{fig:hamconv}
\end{figure}
\end{center}

\begin{center} \begin{figure}
\includegraphics*[scale=0.35,angle=-90]{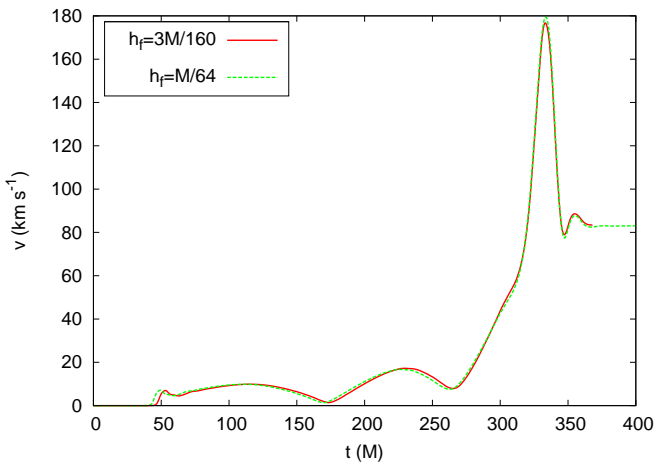}
\caption{The total radiated momentum for the case of unequal masses with
prograde spins (${\rm NE}_{++}$) at two different resolutions.  The
curves have been shifted slightly in time to line up the peaks, for
comparison of the merger dynamics.  The agreement is better than $5\%$
for most of the evolution and better than $2\%$ through the merger and
final kick ($t>300M$).}
\label{fig:kickconv}
\end{figure}
\end{center}

The grid spacing in the finest refinement region, around each black
hole, was $h_f=3M/160$, with the exception of our nonspinning
unequal-mass case, which was one of a set of runs described previously
\citep{Baker06}, and for which we used $h_f=M/40$.  The extraction
radius was at $R=45M$ in every case except for the nonspinning case,
where it was at $R=50M$.  For one of our new physical configurations
(NE$_{++}$) we also extracted at $R=40M$, finding a final kick within
$0.8 {\rm km s^{-1}}$ of that extracted at $R=45M$.  Assuming a radially
dependent error that falls off as $1/R$, as found in a similar kick
computation \citep{Gonzalez07}, this implies that the kick extracted
at $R=45M$ is within $8\%$ of what would be computed at infinite
radius.  Also for this physical configuration we ran a higher
resolution, $h_f=M/64$, to verify that the lower resolution of
$h_f=3M/160$ would be sufficient.  We found satisfactory convergence
of the Hamiltonian constraint (Fig.\ref{fig:hamconv}) and consistency
of the radiated momentum (Fig.\ref{fig:kickconv}).

The thrust $dP^i/dt$ imparted by the radiation was derived
by \cite{Newman:1981fn}, and is computed
from a surface integral of the squared time-integral of the 
radiative Weyl scalar $\psi_4$ times the unit radial vector, 
as given explicitly by \cite{Campanelli:1998jv}:
\begin{eqnarray}
\frac{dP^i}{dt} &=& \lim_{R \rightarrow \infty} \left\{ \frac{R^2}{4 \pi} \oint d\Omega \frac{x^i}{R} \left| \int^t_{-\infty} dt \, \psi_4 \right|^2
\right\}. \label{eqn:thrust_formula}
\end{eqnarray}
To perform the angular integration in (\ref{eqn:thrust_formula}), we
use the second-order Misner algorithm described in
\cite{Misner:1999ab} and \cite{Fiske:2005fx}.  This result is then
integrated numerically to give the total radiated momentum $P^i$; to
obtain the final velocity of the merged remnant black hole, we divide
$P^i$ by the final black-hole mass, as computed from the difference of
the initial ADM mass and the total radiated energy.

\section{Results and Discussion}

\begin{center}
\begin{figure}
\includegraphics*[scale=0.35,angle=0]{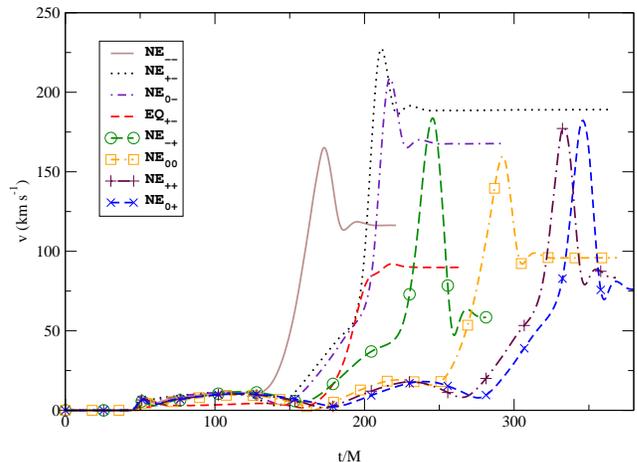}
\caption{Aggregated recoil kicks from all runs listed in Table
  \ref{tab:initial}. The merger time for each binary matches the peak
  in its kick profile; the relative delay in merger times between data
  sets differing in initial spins is consistent with the results of
  \cite{Campanelli:2006uy}.  All configurations show a marked
  "un-kick" after the peak, with the exception of the equal-mass case,
  ${\rm EQ}_{+-}$.}
\label{fig:all_kicks}
\end{figure}
\end{center}

In Fig. \ref{fig:all_kicks} we present the aggregated recoil kick from
each of our simulations.  The kicks obtained range from $\sim 60 {\rm
km s^{-1}}$, in the case where the larger hole's spin is aligned, and the
smaller anti-aligned, with the orbital angular momentum, to $\sim 190
{\rm km s^{-1}}$, when the alignments are reversed.  The kicks possess a
common profile, with the bulk of the momentum radiated over $\sim
50\,M$ before merger. In all but the equal-mass case (${\rm
EQ}_{+-}$), we observe a sharp monotonic rise in kick over $40\,M$,
followed by a substantial ``un-kick''. That is, around the time of
merger and ringdown, we often observe a sudden thrust in momentum that
is directed counter to the momentum that had accumulated during
inspiral.  In the ${\rm EQ}_{+-}$ case, this un-kick is absent.  We
summarize the final kicks for each of our configurations in the first
eight rows of Table~\ref{tab:results}.

It has been noted by \cite{Campanelli:2006uy} that the presence of
spins on black holes in an inspiraling binary can significantly extend
or reduce the time to merger, depending on whether the spins are
aligned or anti-aligned with the orbital angular momentum. We observe
a similar trend in merger times, as illustrated in the peak of the
aggregated recoil kicks.  This tendency had been also been expected
based on PN calculations which show that the last stable orbit is
pushed to smaller radius, implying later merger, for aligned spins
\citep{Damour:2001tu}.  Note that, although resolution can also affect
merger time, for these short runs we have sufficiently resolved the 
dynamics that numerical error in merger time appears negligible
compared to the effect of spin, as demonstrated in Fig.~\ref{fig:kickconv}.

Our simulation results, together with data reported by other groups,
allow us to consider a simple description of the total kick for
arbitrary mass ratio when the two black holes spins are aligned or
anti-aligned with the system's orbital angular momentum.
 
Several recent papers \citep{Herrmann07,Koppitz07,Choi07} have
suggested that the kick velocity resulting from comparable-mass binary
black hole mergers may be approximately described in terms of a simple
scaling dependence consistent with the scaling in the leading-order
post-Newtonian approximation treatment \citep{Kidder95a}.  For kicks
generated by non-spinning black holes of unequal masses, this produces
the Fitchett scaling, which provides a reasonable approximation to
recent numerical simulation results \citep{Gonzalez06}. Recent papers
on spinning black hole mergers suggest that the kicks from nearly
equal-mass mergers may scale linearly, through the quantity
$\Delta=q\hat{a}_1-\hat{a}_2$, with spins that are either aligned or
anti-aligned with the orbital angular momentum
\citep{Herrmann07,Koppitz07}.  For head-on collisions, \cite{Choi07}
have shown that the PN expressions describe the scaling of the
spin-asymmetry kick and its relation to the kick induced by
mass-asymmetry, correctly predicting the relative directions of the
two kick contributions and supporting the idea that the effects of
asymmetries in spins and masses can be considered independently.

For inspiraling mergers, with the assumption that the radial velocity
is small compared to the tangential velocity, the PN prediction for
the cases we consider is that the instantaneous thrust generated by
the spin-asymmetry will be aligned with that of the mass-asymmetry,
though the relative size of these effects may vary as the merger
proceeds.  Unlike the head-on collision case, the direction of thrust
should vary as the system revolves in inspiraling mergers, so that we
cannot infer that the net effects of spin- and mass-asymmetry will
produce collinear contributions to the overall kick.  Motivated by
these observations we will consider our data with the hypothesis that
the magnitudes of the kicks induced by spin- and mass-asymmetries each
scale independently with the PN-predicted scaling, but that the
directional alignment of these two contributions to the kick may
differ by some angle $\theta$.  The total kick may then be of the form
\begin{eqnarray}
v& = &V_0[32\,q^2/(1+q)^5] \nonumber \\
&& \times \sqrt{(1-q)^2 + 2\,(1-q)\,K\,\cos\theta + K^2 },
\label{eqn:kick_formula}
\end{eqnarray}
where $K = k\,(q\hat{a}_1-\hat{a}_2)$. The parameter $V_0$ gives the
overall scaling of the kick (note that the factor in brackets becomes
unity for $q=1$), while $k$ gives the relative scaling of the kick
contributions from spin- and mass-asymmetries.  This expression amounts
to a generalization of the post-Newtonian-inspired kick formula
discussed by \cite{FHH04} and \cite{Koppitz07}, which was consistent
with collinearity of the two kick contributions, $\theta=0$.

We have tested the simple formula (\ref{eqn:kick_formula}) with our
own kick speeds, together with published kicks from \cite{Koppitz07}
and \cite{Herrmann07}, for a total of 17 independent data points.
Without more precise knowledge of the uncertainties for each
measurement it is not possible to do a true statistical fit.  As a
substitute, however, we have obtained values for the three free
parameters $V_0$, $\theta$, and $k$ by minimizing the total of a
$\chi^2$-like quantity $\propto\sum_i (v_{\rm
pred,i}-v_{\rm num,i})^2/v_{\rm num,i}$, where $v_{\rm pred,i}$ is the
predicted, and $v_{\rm num,i}$ is the measured, kick speed for the
$i$th combination of parameters.  The proportionality constant is
chosen so that the minimum of ``$\chi^2$'' is 14, equal to the number
of degrees of freedom.

With this procedure, our best fit to all simulations currently
reported gives $V_0=276 {\rm km~s^{-1}}$, $\theta=0.58$ rad, and
$k=0.85$.  The minimal regions containing 95\% of the probability for
each parameter are $V_0=267-294 {\rm km~s^{-1}}$, $\theta=0.45-0.65$
rad, and $k=0.8-0.89$.  In Table~\ref{tab:results} we compare the
predictions of this formula to our results and those of other groups
who have explored prograde or retrograde spins.  We see that this
simple formula performs well, with errors less than $\sim 10\%$ in all
cases.

We can estimate the uncertainty in $\theta$, and judge how strongly we
can rule out a constant $\theta=0$, by making the conservative
assumption that all the numerical kick results have 10\% statistical
errors.  A chi-squared analysis then indicates that at one standard
deviation $\theta=0.58\pm0.8$ rad.  We also find that, formally,
$\theta=0$ is ruled out at $>5\sigma$ and $\theta=\pi/2$ (sum in
quadrature) is ruled out at $>12\sigma$, but this far from our best
values the error contours are clearly non-Gaussian. It is possible
that a more complicated model (e.g., one in which $\theta$ depends on
the mass ratio) is a better representation than $\theta$=constant, but
no motivation for this exists in the current data.

\begin{deluxetable}{lrrrrrr}
\tablecolumns{7}
\tablewidth{0pt}
\tablecaption{Predicted versus computed kick speed. 
Runs labeled ``S0.\#\#'' are taken from \cite{Herrmann07}, while runs
labeled ``r\#'' are taken from \cite{Koppitz07}. \label{tab:results}}
\tablehead{\colhead{Run} & \colhead{$q$} & \colhead{$\hat{a}_1$} & \colhead{$\hat{a}_2$} & \colhead{$v_{num}$} & \colhead{$v_{pred}$} &
\colhead{$\frac{|\Delta v|}{v_{num}}$(\%)}}
\startdata
 ${\rm NE}_{--}$ & 0.654 &-0.201 &-0.194 &116.3 &119.5 & 2.7\\
 ${\rm NE}_{-+}$ & 0.653 &-0.201 & 0.193 & 58.5 & 58.2 & 0.5\\
 ${\rm NE}_{0-}$ & 0.645 & 0.000 &-0.195 &167.7 &153.1 & 8.7\\
 ${\rm NE}_{00}$ & 0.677 & 0.000 & 0.000 & 95.8 & 98.6 & 2.9\\
 ${\rm NE}_{0+}$ & 0.645 & 0.000 & 0.194 & 76.9 & 71.7 & 6.8\\
 ${\rm NE}_{+-}$ & 0.655 & 0.201 &-0.194 &188.6 &181.9 & 3.6\\
 ${\rm NE}_{++}$ & 0.654 & 0.201 & 0.194 & 83.4 & 92.4 &10.8\\
\medskip
 ${\rm EQ}_{+-}$ & 1.001 & 0.198 &-0.198 & 89.8 & 92.6 & 3.2\\
S0.05 & 1.000 & 0.200 &-0.200 & 96.0 & 93.8 & 2.3\\
S0.10 & 1.000 & 0.400 &-0.400 &190.0 &187.6 & 1.2\\
S0.15 & 1.000 & 0.600 &-0.600 &285.0 &281.5 & 1.2\\
\medskip
S0.20 & 1.000 & 0.800 &-0.800 &392.0 &375.3 & 4.3\\
   r0 & 1.000 &-0.584 & 0.584 &260.0 &274.0 & 5.4\\
   r1 & 0.917 &-0.438 & 0.584 &220.0 &220.8 & 0.3\\
   r2 & 0.872 &-0.292 & 0.584 &190.0 &178.1 & 6.3\\
   r3 & 0.848 &-0.146 & 0.584 &140.0 &141.9 & 1.4\\
   r4 & 0.841 & 0.000 & 0.584 &105.0 &110.4 & 5.1\\
\enddata
\end{deluxetable}

The success of our fit in describing the existing data suggests 
that this simple expression may, for many astrophysical simulations,
adequately describe the dependence of the kicks on the portion of
mass-ratio and spin parameter space that we have studied.  However, 
recent simulations have suggested that the dominant component 
of the kick may be out of the orbital plane
deriving from spins which lie in the orbital plane \citep{Campanelli07}, 
a configuration outside the parameter space we have studied.
\cite{Gonzalez07} have shown that such a configuration can produce 
kicks which may exceed 2500~km~s$^{-1}$ directed out of the orbital
plane.  These results apparently confirm the early predictions of
\cite{RR89} that out-of-plane kicks would be particularly significant.

Since initial submission of this paper, \cite{Campanelli07} have
suggested a combined formula, which generalizes our formula
(\ref{eqn:kick_formula}) to include out-of-plane kicks ($v_{||}$ in
their notation) of the kind discussed above.  If kick speeds $>1000
{\rm km~s^{-1}}$ are common in comparable-mass mergers of black holes
with substantial spin, this comes into apparent conflict with the
observation that essentially all galaxies with bulges appear to have
supermassive black holes in their cores, since galactic escape speeds
tend to be $<1000 {\rm km~s^{-1}}$ (see \citealt{FF05} for a review of
supermassive black holes and their correlation with galactic
properties).  It seems unlikely that most supermassive black holes
have low enough spins to guarantee small kicks, given evidence such as
broad Fe K$\alpha$ lines from a number of black holes
\citep{Iwasawa96,Fab02,Miller02,RN03,RBG05,BR06} as well as overall
arguments from the inferred high average radiation efficiency of
supermassive black holes \citep{Sol82,YT02}. \cite{Bogdanovic:2007hp}
suggest that torques from gas-rich mergers tend to align the spins of
the holes, and thus lead to small kicks, but no known preferred
alignment exists for gas-poor mergers. More exploration of the
parameter space of spin kicks is clearly necessary.

\acknowledgments

This work was supported in part by NASA grant 06-BEFS06-19 and NSF
grant AST 06-07428.  The simulations were carried out using Project
Columbia at the NASA Advanced Supercomputing Division (Ames Research
Center) and at the NASA Center for Computational Sciences (Goddard
Space Flight Center). B.J.K. was supported by the NASA Postdoctoral
Program at the Oak Ridge Associated Universities.  S.T.M. was
supported in part by the Leon A. Herreid Graduate Fellowship.

\end{document}